\documentclass[aps,prl,twocolumn]{revtex4}
\usepackage{graphicx}
\usepackage{amsmath}
\usepackage{amsfonts}

\begin{document}


\title{Attosecond delays in molecular photoionization}


\author{Martin Huppert}
\affiliation{Laboratorium f\"ur Physikalische Chemie, ETH Z\"urich,\\
Vladimir-Prelog-Weg 2, 8093 Z\"urich, Switzerland}
\author{Inga Jordan}
\affiliation{Laboratorium f\"ur Physikalische Chemie, ETH Z\"urich,\\
Vladimir-Prelog-Weg 2, 8093 Z\"urich, Switzerland}
\author{Denitsa Baykusheva}
\affiliation{Laboratorium f\"ur Physikalische Chemie, ETH Z\"urich,\\
Vladimir-Prelog-Weg 2, 8093 Z\"urich, Switzerland}
\author{Aaron von Conta}
\affiliation{Laboratorium f\"ur Physikalische Chemie, ETH Z\"urich,\\
Vladimir-Prelog-Weg 2, 8093 Z\"urich, Switzerland}
\author{Hans Jakob W\"orner}
\email[]{woerner@phys.chem.ethz.ch}
\homepage[]{www.atto.ethz.ch}
\affiliation{Laboratorium f\"ur Physikalische Chemie, ETH Z\"urich,\\
Vladimir-Prelog-Weg 2, 8093 Z\"urich, Switzerland}


\date{\today}

\begin{abstract}
We report measurements of energy-dependent attosecond photoionization delays between the two outer-most valence shells of N$_2$O and H$_2$O. The combination of single-shot signal referencing with the use of different metal foils to filter the attosecond pulse train enables us to extract delays from congested spectra. Remarkably large delays up to 160 as are observed in N$_2$O, whereas the delays in H$_2$O are all smaller than 50 as in the photon-energy range of 20-40 eV. These results are interpreted by developing a theory of molecular photoionization delays. The long delays measured in N$_2$O are shown to reflect the population of molecular shape resonances that trap the photoelectron for a duration of up to $\sim$110 as. The unstructured continua of H$_2$O result in much smaller delays at the same photon energies. Our experimental and theoretical methods make the study of molecular attosecond photoionization dynamics accessible.
\end{abstract}


\maketitle

Photoionization and photoelectron spectroscopies are powerful approaches to measuring the electronic structure of matter \cite{berkowitz79a,kimura81a}. A complete quantum-mechanical description of photoionization, both in the time and frequency domains, requires the amplitude and phase of all dipole matrix elements, e.g. in a partial-wave expansion. Most experiments to date measure photoionization cross sections which are described by the sum of squared moduli of individual partial-wave dipole matrix elements. Cross sections thus contain no information about the partial-wave phase shifts. In contrast, photoelectron angular distributions are highly sensitive to partial-wave phase shifts \cite{cooper68a,seideman02a} between continua associated with the same ionic state. Phase-shifts between continua associated with different ionization thresholds are not measurable in the frequency domain because the lack of spectral overlap between the corresponding photoelectrons erases the coherence required to measure such phase shifts.

In this letter, we show that attosecond metrology can be employed to measure this information in molecular photoionization. Specifically, we study the effect of molecular shape resonances on the measured photoionization delays. When the combined molecular (or atomic) and centrifugal potential felt by the photoelectron displays a barrier, one or several quasi-bound states can emerge \cite{dehmer72a,dehmer79a,starace81a}. These resonances decay by tunneling through the potential barrier and often lead to a local enhancement of the photoionization cross section. Such shape resonances have so far only been measured by frequency-resolved measurements. Here, we show that attosecond metrology provides access to the time-domain manifestation of shape resonances. In the case of N$_2$O, our measurements indeed reveal surprisingly large delays reaching up to 160 as in the range of 20 to 40 eV. In contrast, delays measured at the same photon energies in H$_2$O all lie below 50 as in magnitude. These results are interpreted by developing a theory of molecular photoionization delays relying on accurate molecular scattering calculations. This analysis shows that the delays measured in the case of N$_2$O are caused by the population of molecular shape resonances that decay by tunneling on a time scale of up to $\sim$ 110 as. The small delays measured in H$_2$O are shown to reflect the flat nature of the associated photoionization continua.

Our work builds upon recent pioneering experiments which have demonstrated the possibility of measuring delays in the photoionization of atoms and atomic solids \cite{cavalieri07a,schultze10a,kluender11a,locher15a}, which were comprehensively reviewed in Ref. \cite{pazourek15a}. These experiments have been recognized to be sensitive to the scattering phase, or, more precisely, to its energy derivative, known as the Wigner-Smith delay \cite{wigner55a,smith60a}. Transposing these methods to molecules, both in the gas and condensed phases, has remained challenging so far because of the associated spectral congestion. The method of attosecond streaking \cite{schultze10a} is difficult to apply to the molecular valence shell because most molecules feature multiple shells within the bandwidth of a typical attosecond pulse. The method of attosecond interferometry using an extreme-ultraviolet (XUV) attosecond pulse train (APT) and a synchronized infrared (IR) pulse, which was successfully used for measuring atomic photoionization delays \cite{kluender11a}, offers higher spectral resolution as a consequence of the quasi-discrete nature of the XUV spectrum, without sacrificing temporal resolution. Spectral overlap of photoelectron spectra associated with different harmonic orders is the principle remaining limitation of this method and is addressed in our work. An important step towards extending such measurements to molecules was taken in Ref. \cite{haessler09a}, which did however not report photoionization delays, but focused on the role of Rydberg states in attosecond interferometry.

The experimental setup consists of an actively-stabilized attosecond beamline \cite{huppert15a} and a magnetic-bottle photoelectron spectrometer \cite{jordan15a}. High-harmonic generation in a gas cell filled with 10 mbar argon is driven by a 1.5 mJ, 30-fs laser pulse centered at 800 nm. These laser pulses are generated by an amplified titanium:sapphire laser system (Femtopower Pro V CEP by Femtolasers). The generated attosecond XUV pulse train is separated from the residual IR pulse under vacuum by means of a perforated off-axis parabolic mirror that simultaneously recollimates the IR beam. The XUV pulse train is then spectrally filtered by 100-nm thick metal foils. The tin filter transmits harmonic orders 11-15 (16.4-23.9 eV), the titanium filter the orders 15-21 (26.5-32.5 eV) and the chromium filter the orders 23-27 (35.3-43.4 eV). The XUV beam is refocused by a toroidal mirror and is recombined with the IR beam by means of a second perforated off-axis parabolic mirror. The delay between XUV and IR is varied by tuning the length of the IR beam path and is actively stabilized during the measurement to a residual jitter of $\sim$ 30 as \cite{huppert15a}. Photoelectrons generated from the overlapping XUV and IR pulses are collected and energy analyzed using a $\sim$0.9~m-long magnetic-bottle spectrometer equipped with a permanent magnet. Electron time-of-flight spectra are acquired on a single-shot basis using a digitizer card. A chopper wheel is used in the IR beam path to block every other laser shot and to record two-color (XUV/IR) and one-color (XUV-only) spectra in immediate temporal sequence. This approach substantially improves the signal-to-noise ratio and generates high-fidelity difference spectrograms that are used in the subsequent analysis.

\begin{figure}[h!]
\includegraphics[width=0.5\textwidth]{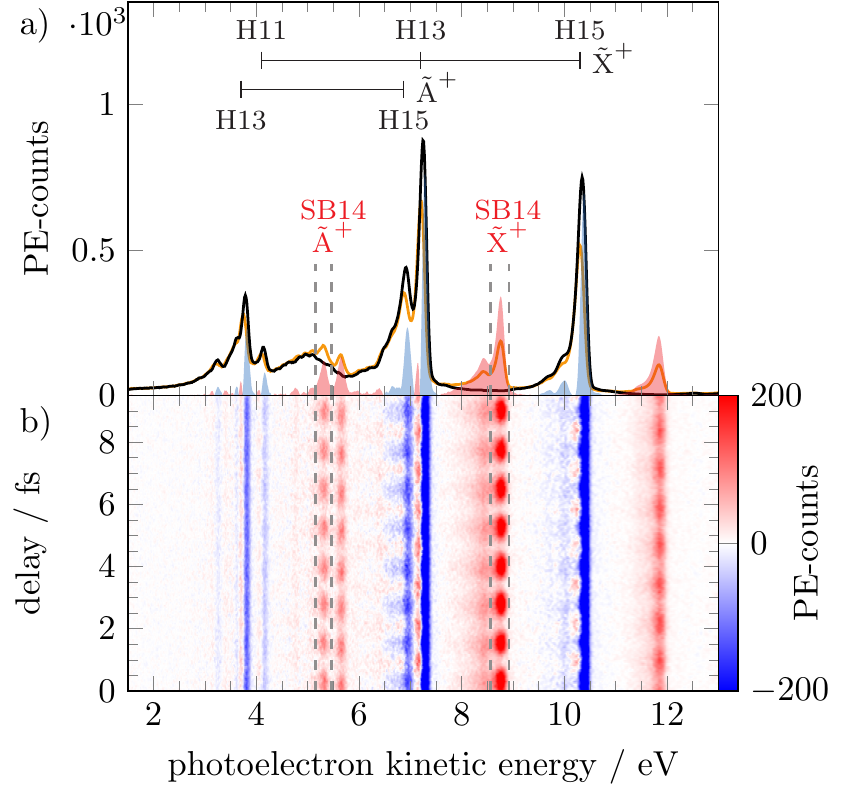}
\caption{(color online) a) Photoelectron spectrum of N$_2$O generated by an attosecond pulse train transmitted through a Sn filter (black line) and in the presence of the dressing IR field (orange line). Difference spectra, obtained by subtracting XUV-only from XUV/IR photoelectron spectra and vice-versa, are shown in red and blue, respectively. b) Difference spectrum as a function of the IR-XUV delay.}
\end{figure}

Figure 1a shows the photoelectron spectrum of N$_2$O recorded following ionization by a Sn-filtered attosecond pulse train (black line). The spectrum is dominated by photoelectrons associated with the two lowest-lying electronic states, $\tilde{\rm X}^+$ and $\tilde{\rm A}^+$ of N$_2$O$^+$ and the harmonic orders 11, 13 or 15. These two electronic states are both bound and have vertical ionization potentials of 12.89 and 16.38~eV, respectively \cite{kimura81a}. Difference spectra obtained by subtracting one-color (XUV-only) from two-color (XUV/IR) spectra feature both positive contributions (red) resulting from two-photon ionization and negative contributions (blue) reflecting the depletion of the one-color XUV signal. 

Figure 1b shows the difference spectrogram acquired by averaging $\sim$ 5000 laser shots per delay. Three side bands display pronounced modulations and are assigned, in order of increasing kinetic energy, to side band (SB) 14 of $\tilde{\rm A}^+$, SB12 of $\tilde{\rm X}^+$ (not labeled) and SB14 of $\tilde{\rm X}^+$. 
The clean separation of the individual side bands in the difference spectrogram allows for a straightforward determination of the phase shift between SB14 of $\tilde{\rm A}^+$ and SB14 of $\tilde{\rm X}^+$, i.e. the determination of the attosecond photoionization delays between these two electronic states. We find a delay of $\tau(\tilde{\rm A}^+)-\tau(\tilde{\rm X}^+)$ = 35$\pm 6$ as. All delays were determined by Fourier transforming the delay axis and evaluating the phase as a complex average over the frequency of the side-band oscillation. The results of individual measurements are weighted with the root-mean-square phase noise over the integrated electron kinetic-energy range and averaged over several measurements (see Table S1 for details) to yield the final delays and the associated 95 \% confidence intervals. We note that SB16 of $\tilde{\rm X}^+$, located close to 12 eV kinetic energy, oscillates nearly out of phase to SB14 of $\tilde{\rm X}^+$. This is the case because the strong suppression of H17 prevents the two-pathway interference to take place, such that the modulation of SB16 is dominated by the periodic depletion of the H15 $\tilde{\rm X}^+$ photoelectron band.

We now show that our approach is more widely applicable by performing similar measurements on H$_2$O. The photoelectron spectra are dominated by bands associated with the $\tilde{\rm X}^+$ and $\tilde{\rm A}^+$ states of H$_2$O$^+$ generated by the harmonic orders 11, 13 or 15. These two electronic states are both bound and have vertical ionization potentials of 12.62 and 14.74 eV, respectively \cite{kimura81a}. The XUV/IR spectrum (Fig. 2a, orange line) displays significant spectral overlap between two-color and one-color signals.  

\begin{figure}[h!]
\includegraphics[width=0.5\textwidth]{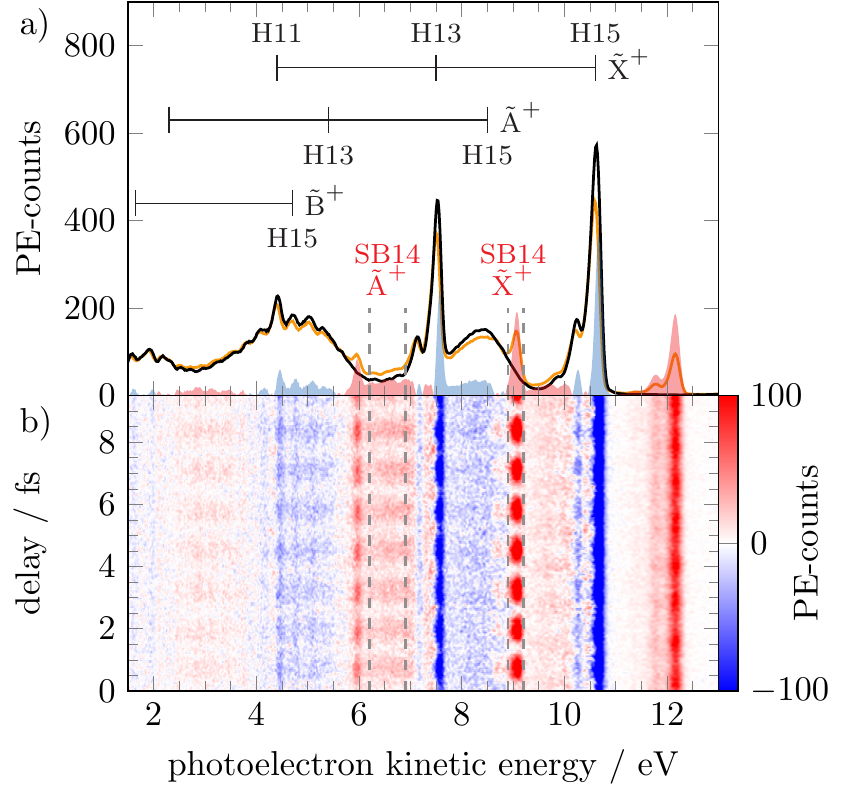}
\caption{(color online) Same as Fig. 1 for H$_2$O.}
\end{figure}

In spite of the spectral overlap, the difference spectrum (red and blue lines) again displays a sufficient spectral separation for identifying the contributions from individual harmonic orders and electronic states of the cation. The regions corresponding to SB14 of $\tilde{\rm A}^+$ and $\tilde{\rm X}^+$ are again easily identified. Integrating over the ranges indicated in Fig. 2, we obtain a photoionization delay of $\tau(\tilde{\rm A}^+)-\tau(\tilde{\rm X}^+)$ = 10$\pm 5$ as.

Similar measurements have been repeated with titanium and chromium foils and all results are summarized as blue squares in Figs. 3a and 3b. The delays measured in N$_2$O increase from 21.7 eV to 31.0 eV, where a comparably large delay of 160$\pm$34 as is reached. In contrast, the delays measured in H$_2$O are much smaller. These experimental results are now compared to calculations, enabling us to interpret the delays and assign them to particular structures of the photoionization continua.

The theory of atomic photoionization delays has recently received much attention \cite{kheifets10a,nagele11a,pazourek12a,dahlstrom12a,dahlstrom13a,jimenez14a,gaillac16a}. In contrast, a theory for molecular photoionization delays has not been developed so far, although calculations of such delays have been performed for H$_2$ and H$_2^+$, taking advantage of the analytical form of two-center Coulomb functions \cite{kawai07a,ivanov12a,serov13a,ning14a}. Very interesting insight has also been obtained from one-dimensional model systems \cite{caillat11a,chacon14a}. 

In contrast to atoms, molecules are not spherically symmetric. Additional complexity originates from the dependence of the photoionization matrix elements on the orientation of the molecule. We therefore first derive the expressions of molecular photoionization delays in the molecular frame (MF) and subsequently introduce the averaging over emission angles and molecular orientations to predict delays measured in the laboratory frame (LF). Details of the theory will be given elsewhere \cite{baykusheva16b}. 

In a single-center expansion, the length-form dipole matrix elements for photoionization by linearly polarized radiation are given by
\begin{equation}
\langle\Psi_{f,{\bf{\kappa}}}^{(-)}|{\bf r}\cdot\hat{n}|\Psi_i\rangle = \sqrt{\frac{4\pi}{3\kappa}}\sum_{\ell m\nu} I_{\ell m \nu} Y_{\ell m}({\hat{\kappa}})Y^*_{1\nu}({\hat n}),
\end{equation}
where $|\Psi_i\rangle$ is the initial bound state, ${\bf r}$ is the position operator, $Y_{\ell m}$ are the spherical harmonics, $Y_{1,\nu=0,\pm 1}$ expresses the dependence on the orientation $\hat{n}$ of the XUV polarization in the MF and $|\Psi_{f,{\bf{\kappa}}}^{(-)}\rangle$ denotes the anti-symmetrized product of the ionic and photoelectron wave functions of asymptotic momentum $\hat{\kappa}$. $I_{\ell m \nu} = \sqrt{2/\pi}(-i)^{\ell}\langle{\Psi^{(-)}_{f,\kappa \ell m}}|{\bf r}_{\nu}|\Psi_i\rangle$ is the partial-wave matrix element. 

These matrix elements can be used to define an effective Wigner-Smith delay in the MF depending on the photon energy $E$ and the angles $\hat{\kappa}$ and $\hat{n}$:
\begin{equation}
\tau^{\rm MF}_{\rm WS}(E,\hat{\kappa},\hat{n})=\hbar\frac{\partial}{\partial E}\left[\arg\left(\langle\Psi_{f,{\bf{\kappa}}}^{(-)}|{\bf r}\cdot\hat{n}|\Psi_i\rangle\right)\right].
\end{equation}

Our experiment does not directly measure this quantity, but rather a related delay defined by two-photon XUV/IR transitions. The corresponding two-photon matrix elements neglecting bound-state contributions can be derived on the basis of the asymptotic approximation also used in \cite{dahlstrom13a}:
\begin{equation}
M(\hat{k},\kappa,\hat{R}) \approx C_{k,\kappa}\sum_{\substack{L=\ell\pm 1\\M=m+\nu'}} Y_{LM}({\hat{ k}})b_{LM}(\hat{R}),
\label{2pmel}
\end{equation}
where   
\begin{equation}
b_{LM}(\hat{R})=\sum_{\ell m\nu'\nu}\langle Y_{LM}|Y_{1\nu'}|Y_{\ell m}\rangle Y^*_{1\nu'}(\hat{R})I_{\ell m\nu}Y^*_{1\nu}(\hat{R}),
\end{equation}
$\hat{R}=(\alpha,\beta,\gamma)$ are the Euler angles that rotate the MF to coincide with the LF,
$\nu'=0,\pm 1$ defines the IR transition as being parallel or perpendicular and
\begin{eqnarray}
C_{k,\kappa} &=& -\frac{2\pi^2}{3}(8\pi)^{3/2}E_{\omega}E_{\Omega}N_kN_{\kappa} \\ \nonumber
			&\times & \frac{1}{|k-\kappa|^2}\exp{\left(-\frac{\pi Z}{2}\left(\kappa^{-1}-k^{-1}\right)\right)}  \\ \nonumber
			&\times & \frac{(2\kappa)^{iZ/\kappa}}{(2k)^{iZ/k}}\frac{\Gamma(2+iZ(\kappa^{-1}-k^{-1}))}{(\kappa-k)^{iZ(\kappa^{-1}-k^{-1})}}.				
\end{eqnarray}
Here, $\kappa$ and $k$ are the asymptotic momenta of the electron following interaction with XUV or XUV/IR, respectively. $E_{\omega}$ and $E_{\Omega}$ are the electric fields of the IR and XUV pulses, respectively, and $Z$ is the charge number of the molecule after photoionization. 

The molecular photoionization delay measured in side band of order $2q~(q\in \mathbb{N})$ is
\begin{multline} \label{2pdelmf}
\tau(2q,\hat{k},\hat{R})=\frac{1}{2\omega}\arg \left[ C_{2q+1}C^*_{2q-1} \right.  \\ 
\left. \times \sum_{LL'MM'}Y^*_{LM}(\hat{k})Y_{L'M'}(\hat{k})b^*_{LM,2q-1}(\hat{R})b_{L'M',2q+1}(\hat{R})\right],
\end{multline}
where $2q\pm 1$ labels the order of the harmonics involved in the one-photon XUV transitions. The relation to $k$ and $\kappa$ is given by $k^2/2=2q\omega-I_{\rm p}$ and $\kappa^2/2=(2q\pm1)\omega-I_{\rm p}$. 

Since our measurements are not resolved in $\hat{k}$, we integrate the expressions leading to Eq. (\ref{2pdelmf}) over $\hat{\kappa}$ to obtain
\begin{equation} \small
\tau(2q,\hat{R})=\frac{1}{2\omega}\arg\left[C_{2q+1}C^*_{2q-1}\sum_{LM}b^*_{LM,2q-1}(\hat{R})b_{LM,2q+1}(\hat{R})\right].
\end{equation}
This derivation shows that molecular photoionization delays can still be interpreted as the sum of a measurement-induced (continuum-continuum) delay $\tau_{\rm cc}(2q)=\arg(C_{2q+1}C^*_{2q-1})$ \cite{kluender11a,dahlstrom12a} and a molecule-specific part of the delay. Since our measurements are done on randomly aligned molecules, we further convolute them over an isotropic axis distribution to obtain the effective delay measured in the laboratory frame 
\begin{equation} \label{lfdel}
\tau_{\mathrm{mol}}(2q)=\frac{1}{2\omega}\arg\left[\int{\mathrm d}\hat{R}\sum_{LM}b_{LM,2q-1}^*(\hat{R})b_{LM,2q+1}(\hat{R})\right].
\end{equation}
The total delays measured in the laboratory frame are thus given by $\tau_{\rm tot}(2q)=\tau_{\mathrm{cc}}(2q)+\tau_{\mathrm{mol}}(2q)$.

\begin{figure}[h!]
\includegraphics[width=0.5\textwidth]{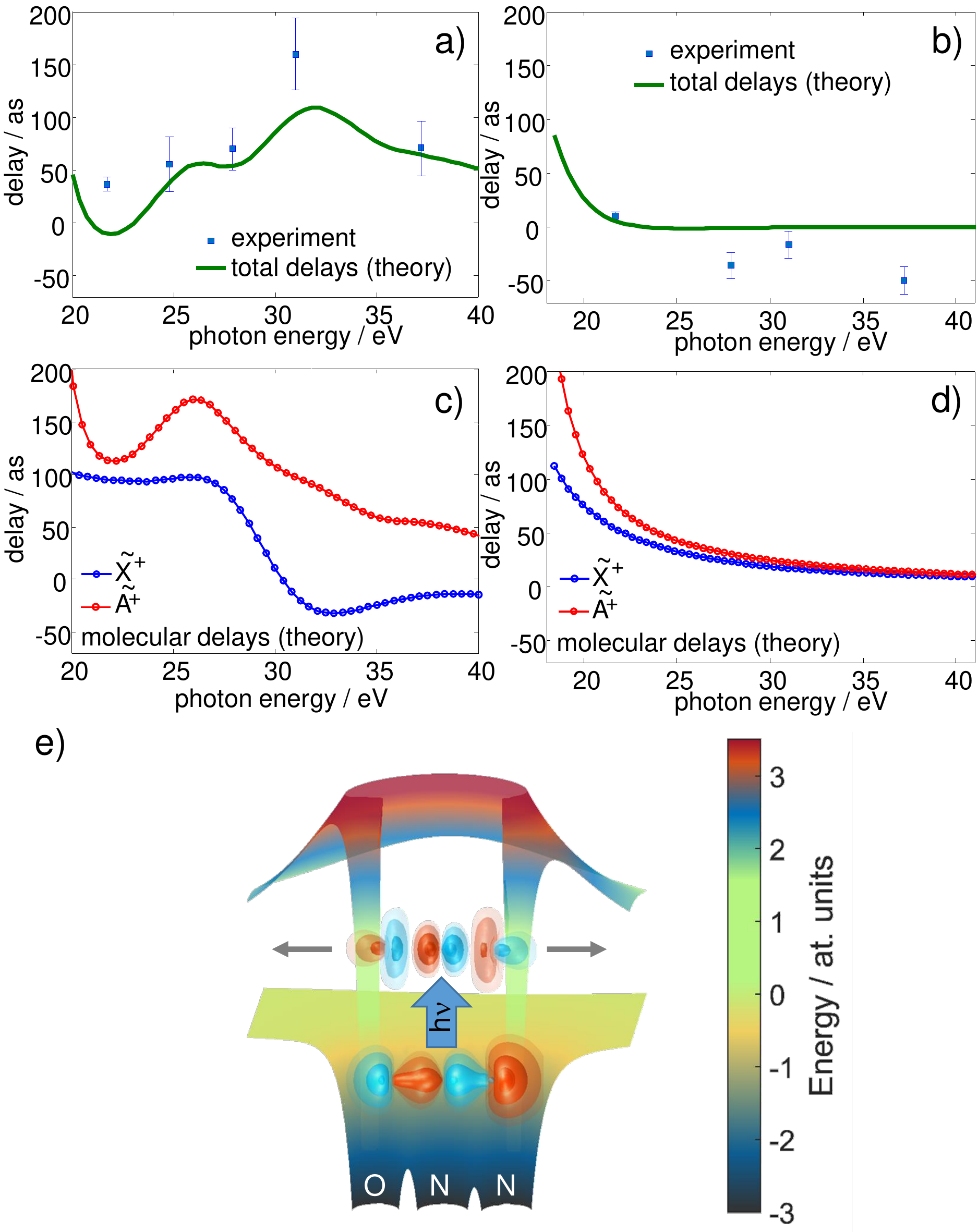}
\caption{(color online) a) Measured and calculated delays between photoelectrons leaving N$_2$O$^+$ in its $\tilde{\rm A}^+$ or $\tilde{\rm X}^+$ states, b) same as a) for H$_2$O, c),d) calculated molecular delays for the two species as defined by Eq. (\ref{lfdel}), e) shape resonance of $\sigma$ symmetry in the photon-energy range of 25-30 eV associated with the $\tilde{\rm A}^+$ state of N$_2$O$^+$. The lower surface shows the numerically calculated molecular potential containing electrostatic and exchange interactions. The upper surface shows the total potential, i.e. the sum of the molecular and centrifugal potentials for $\ell=5$. The wave functions of the bound orbital and the shape-resonant state are illustrated by isosurfaces with color-coded signs. The grey arrows represent tunneling of the photoelectron through the barrier.}
\end{figure}

We now use these results to predict the measured photoionization delays from the output of molecular quantum-scattering calculations. We calculate the partial-wave matrix elements $I_{\ell m\nu}$ of Eq. (1) in the length gauge using the iterative Schwinger variational method implemented in ePolyScat \cite{gianturco94a,natalense99a}. Details about the calculations and comparison with experimental photoionization data are given in the Supplementary Material (Section I, Figs. S1 and S2). 

The difference of the total delays $\tau_{\rm tot}^{\tilde{A}^+}-\tau_{\rm tot}^{\tilde{X}^+}$ (including the continuum-continuum delays) are shown as green lines in Fig. 3, treating the XUV photon energy as a continuous variable and keeping the IR photon energy fixed to the experimental value (1.5498 eV). The agreement with the experimental results (blue squares) is good with all general trends well reproduced. In the case of N$_2$O, the delays are correctly predicted to increase from 21.7 to 31 eV, where they reach a maximum, followed by a decrease. In the case of H$_2$O, the delays are predicted to rapidly converge towards zero in the same energy range. The experimental data support this prediction by lying much closer to zero than in the case of N$_2$O.

Further insight into these results is obtained by studying the molecular photoionization delays associated with specific final ionic states as defined by Eq. (\ref{lfdel}). The investigated range of photon energies (21.7 - 37.2 eV) features two shape resonances, one of $\sigma$ and one of $\pi$ symmetry associated with each of the two lowest electronic states of N$_2$O$^+$. These shape resonances have been identified theoretically in the continua associated with the $\tilde{\rm A}^+$ state \cite{braunstein87a,rathbone05a}. Our calculations confirm these results and additionally reveal the presence of one shape resonance of $\sigma$ and one of $\pi$ symmetry in the $\tilde{\rm X}^+$ continuum range studied. 

Shape resonances occur when the combined centrifugal and molecular potentials support quasi-bound states that decay by tunneling through the potential barrier. We find that the shape resonances of $\sigma$ symmetry have a very pronounced effect on the molecular photoionization delays shown in Fig. 3c. 

We have used the resonance-search algorithm implemented in ePolyScat \cite{natalense99a} to locate these resonances in the photoionization continua of N$_2$O. This algorithm relies on a single-active-electron approximation and is therefore less accurate than the results displayed in Figs. 3a-d, but it provides the approximate location, energy width and wave function of the resonant state. The wave function of the shape resonance of $\sigma$ symmetry associated with the $\tilde{\rm A}^+$ state of N$_2$O$^+$ is shown in Fig. 3e, together with the adiabatic molecular potential and the total (molecular+centrifugal) potential. Our resonance-search calculations predict this shape resonance to lie at 28.8~eV and to have a lifetime of 114 as. The shape resonance of $\sigma$ symmetry associated with the $\tilde{\rm X}^+$ state of N$_2$O$^+$ (not shown) is located at 25.5~eV and has a lifetime of 97~as.

These two shape resonances however have different signatures in the molecular photoionization delays shown in Fig. 3c. The pronounced local maximum in the red curve of Fig. 3c and its width correspond to the expected signature of the shape resonance of $\sigma$ symmetry in the $\tilde{\rm A}^+$ continuum of N$_2$O$^+$, but shifted to lower photon energies compared to the location of the shape resonance. The reasons for this shift will be further discussed in Ref. \cite{baykusheva16b}. The shape resonance causes the outgoing photoelectron to be trapped for the lifetime of the quasi-bound state. Similarly, the local maximum in the photoionization delay associated with the $\tilde{\rm X}^+$ state also corresponds to the location of the $\sigma$-symmetry shape resonance in this channel and the local enhancement is again related to the lifetime of the shape resonance. In this case however, the enhancement of the delay is more asymmetric with a steep drop of the delay towards 30 eV photon energy.

Our measurements thus probe the time-domain manifestation of molecular shape resonances, i.e. the transient trapping of the photoelectron immediately after ionization, followed by its delayed release through tunneling. We note that the lifetime of these resonances could not be extracted from frequency-domain measurements because they do not appear as pronounced local maxima in the partial photoionization cross sections \cite{braunstein87a,rathbone05a}. In fact, the shape resonances in the $\tilde{\rm X}^+$ channel have not been discussed before, neither in the context of experimental nor that of theoretical results.

The remarkable energy dependence of the delays in N$_2$O, caused by the shape resonances, is further highlighted by contrasting these results with the same quantities for H$_2$O (Fig. 3d). In this case, the calculated molecular delays show a smooth decrease as a function of the photon energy, which simply reflects the energy dependence of the Coulomb phase shifts. A very similar behavior is indeed found for photoionization delays of the hydrogen atom (see e.g. Fig. 16 in Ref. \cite{dahlstrom12a}). The absence of shape resonances, Cooper minima \cite{schoun14a} and other continuum structure from the photoionization continuum of H$_2$O in the studied spectral range \cite{machado89a} explains the lack of additional energy dependence of the delays on top of the observed variation. This is largely confirmed by the experimental results. Remaining discrepancies are attributed to approximations made in our state-of-the-art molecular quantum-scattering calculations, i.e. the frozen-core Hartree-Fock approximation and the neglect of channel-interactions in photoionization, as well as the asymptotic approximation made in deriving the two-photon matrix elements in Eq. (\ref{2pmel}).

In conclusion, we have reported the measurement and theoretical interpretation of attosecond delays in the photoionization of molecules. This development makes molecular systems in the gas, liquid \cite{jordan15a} and solid phases accessible to attosecond interferometry. The measured delays are found to be remarkably large in N$_2$O, which is shown to be a consequence of transient trapping of the photoelectron in shape resonances. Our measurements are thus shown to probe new facets of attosecond ionization dynamics. In a broader perspective, they provide the first experimental access to phase shifts between continua associated with different states of the molecular cation, which are not measurable by frequency-domain techniques. This aspect adds considerable value to attosecond metrology. Looking further ahead, such studies can be generalized to obtain complementary information about photoionization-induced dynamics, such as charge migration \cite{cederbaum99a,kraus15b} or intermolecular Coulombic decay in the gas and liquid phases \cite{aziz08a}. 

\begin{acknowledgments}
We acknowledge funding from an ERC Starting Grant (contract 307270-ATTOSCOPE) and the NCCR-MUST, a funding instrument of the Swiss National Science Foundation. We thank S. Hartweg for his contributions to the early experimental phase of this work.

M.~H. and I.~J. contributed equally to this work.
\end{acknowledgments}

\end{document}